\documentclass{moriond}

\graphicspath{{figures/}}

\def\be{\begin{equation}}
\def\ee{\end{equation}}
\def\bea{\begin{eqnarray}}
\def\eea{\end{eqnarray}}

\newcommand{\Photo}{}
\begin{document}
\vspace*{4cm}
\title{An all-sky search in early O3 LIGO data for 
continuous gravitational-wave signals from
unknown neutron stars in binary systems}

\author{Rodrigo Tenorio, for the LIGO Scientific Collaboration \& Virgo Collaboration}
\address{Departament de F\'isica, Institut d'Aplicacions Computacionals i de Codi Comunitari (IAC3), 
Universitat de les Illes Balears, and Institut d'Estudis Espacials de Catalunya (IEEC), 
Carretera de Valldemossa km 7.5, E-07122 Palma, Spain}

\maketitle
\abstracts{
    We present a search for continuous gravitational waves emitted by neutron stars in binary systems 
    conducted on data from the early third observing run of the Advanced LIGO and Advanced Virgo detectors using 
    the semicoherent, GPU-accelerated, BinarySkyHough pipeline. The search analyzes the most sensitive 
    frequency band of the LIGO detectors, 50~--~300 Hz. Binary orbital parameters are split into four 
    regions, comprising orbital periods of 3~--~45 days and projected semimajor axes of 2~--~40 
    light-seconds. No detections are reported. We estimate the sensitivity of the search using simulated 
    continuous wave signals, achieving the most sensitive results to date across the analyzed parameter space.
}

\section{Continuous gravitational waves}
Continuous gravitational waves (CWs) are persistent forms of gravitational radiation. Possible sources include 
rapidly-spinning neutron stars (NS) presenting crustal deformations, r-mode instabilities or free precession.
They are orders of magnitude weaker than compact binary coalescense transient signals \cite{GWTC2}, requiring 
long integration times to be detected. 
Fully-coherent matched filtering searches are unfeasible for wide parameter space surveys 
due to the strong scaling of the number of phase templates to consider with respect to the observing time. 
Instead, semicoherent methods, which allow the use of coarser grids, yield a better sensitivity at a fixed 
computational cost.

We search for CWs from unknown NSs located in circular binary systems. These are quasi-monochromatic signals 
modulated due to the daily and yearly movement of the Earth around the solar system barycenter, on top of 
which there is an extra frequency modulation due to the binary motion of the source. 
Population studies suggest frequency evolution due to the spindown of the source can be 
neglected$\;{}^{2,\;3}$. For a NS in a circular binary orbit, the CW frequency as it arrives 
to an Earth-based detector can be described as
\begin{equation}
    f(t) = f_{0} \cdot 
    \left(1 
    + \frac{\vec{v}(t)}{c} \cdot \hat{n} 
    - a_p \Omega \cos{\left[\Omega (t - t_{\textrm{asc}})\right]} \right)\;,
\end{equation}
where $f_{0}$ is the CW frequency from the source frame at a fiducial reference time, $\hat{n}$ represents 
the sky position of the source, and $a_p, \;\Omega, \;t_{\textrm{asc}}$ are the binary orbital parameters: 
projected semimajor axis, orbital frequency and time of passage through the ascending node, respectively.
$\vec{v}$ is the detector's velocity with respect to the Solar System Barycenter.

Here we summarize the all-sky search for continuous gravitational-wave signals from unknown neutron
stars in binary systems conducted on early O3 Advanced LIGO data using the \texttt{BinarySkyHough} pipeline 
\cite{BSH}. The search and its results were also described in detail in a previous publication \cite{O3a}.

\section{Early O3 LIGO-Virgo data}
The early third observing run of the Advanced LIGO \cite{aligo} and Advanced Virgo \cite{avirgo} detectors 
spans 6 months of data, from 1st April 2019 to 1st October 2019.
We used 1024s-long Short Fourier Transforms of Advanced LIGO data, on which a time-domain cleaning procedure 
was applied to prevent background degradation due to loud, frequent transient noise artifacts~\cite{sfdb}.

\section{Search setup}

\begin{figure}
    \includegraphics[width=\textwidth]{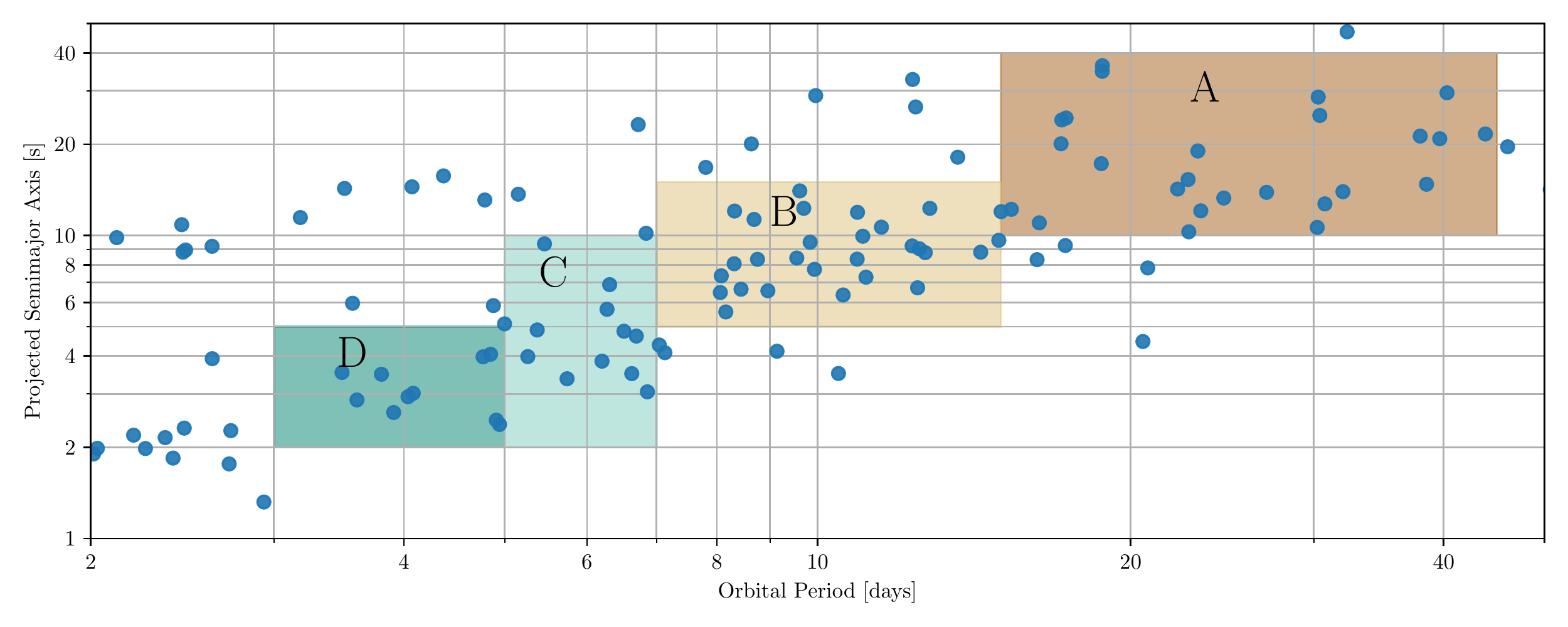}
    \caption{Parameter space of binary circular orbital parameters covered by the search. Time of passage
    through the ascending node templates were placed along the interval 
    $[t_{\textrm{mid}} - P/2,\; t_{\textrm{mid}} + P/2]$ for each orbital parameter $P$.
    The low-frequency search covered regions A, B, C and D, while the high-frequency search was entirely
    focused on region B. Blue dots represent orbital parameters from pulsars in the ATNF 
    catalog~\protect\cite{atnf}. Figure reproduced from~\protect\cite{O3a}.}
    \label{fig:parameter_space}
\end{figure}

We analyzed two main frequency bands, namely low frequency [50, 100] Hz and high frequency [100, 300] Hz,
enclosing the most sensitive frequency band of the Advanced LIGO detectors.
Fig.~\ref{fig:parameter_space} shows the binary parameter space regions surveyed by this search. Regions
A, C, D were analyzed at low frequency, while region B was analyzed at high frequency as well.

The main stage of the search consists of the application of \texttt{BinarySkyHough}~\cite{BSH}, 
a semicoherent GPU-accelerated pipeline able to analyze wide parameter-space regions using the Hough 
transform algorithm. The search is partitioned into $0.125\;$Hz sub-bands. The most significant outliers 
of each band are clustered according to their frequency evolution~\cite{ttd}, reducing the effective number 
of outliers to follow up.

We collect the loudest outlier from each of the top 5 clusters of each sub-band. A first veto discards those
outliers whose frequency evolution crosses an identified narrow spectral artifact. The surviving outliers
are analyzed using an MCMC-based $\mathcal{F}$--statistic sampler \cite{MCMC} implemented in the 
\texttt{PyFstat} Python package \cite{pyfstat}. This sampler re-analyzes the outliers using a longer coherence
time, imposing a further constraint on its phase evolution. None of the re-analyzed outliers is consistent with
an astrophysical source and we proceed to assess the sensitivity of the search.

\section{Results}

\begin{figure}
    \includegraphics[width=\textwidth]{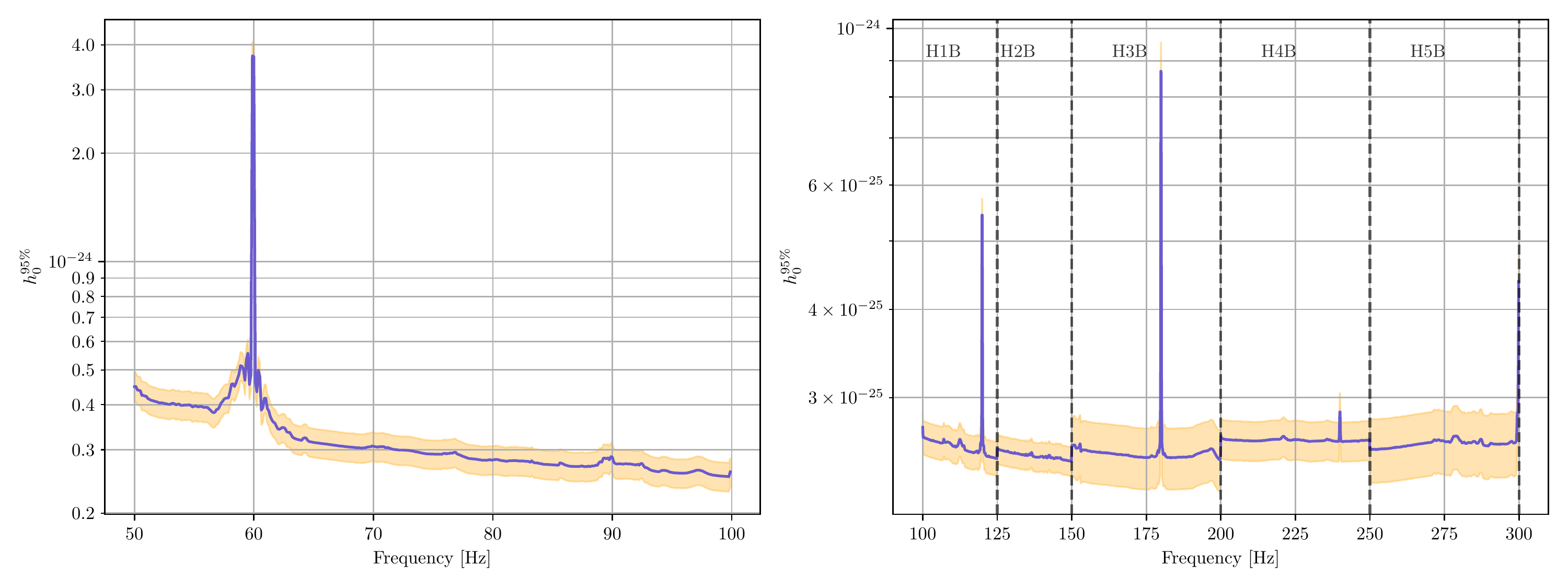}
    \caption{Estimation of the sensitivity achieved by the search in each of the frequency bands.
    These amplitudes correspond to an average $95\%$ detection efficiency assuming a population of 
    isotropically oriented NS.
    For the sake of simplicity, low frequency results are only shown for region B; the other sections
    are within a consistent level of sensitivity.
    Figure reproduced from~\protect\cite{O3a}.}
    \label{fig:UL}
\end{figure}

Search sensitivity is estimated via Monte Carlo studies on a population of simulated signals added to the data.
The amplitude sensitivity, shown in Fig.~\ref{fig:UL}, attains a minimum value of 
\mbox{$h_{0}^{95\%} = (2.4 \pm 0.1) \times 10^{-25}$} at $149.5\;\textrm{Hz}$, improving the results from a 
search on O2 Advanced LIGO data \cite{O2BSH} by a factor \mbox{of $\sim1.6$}.

\begin{figure}
    \centering
    \includegraphics[width=0.5\textwidth]{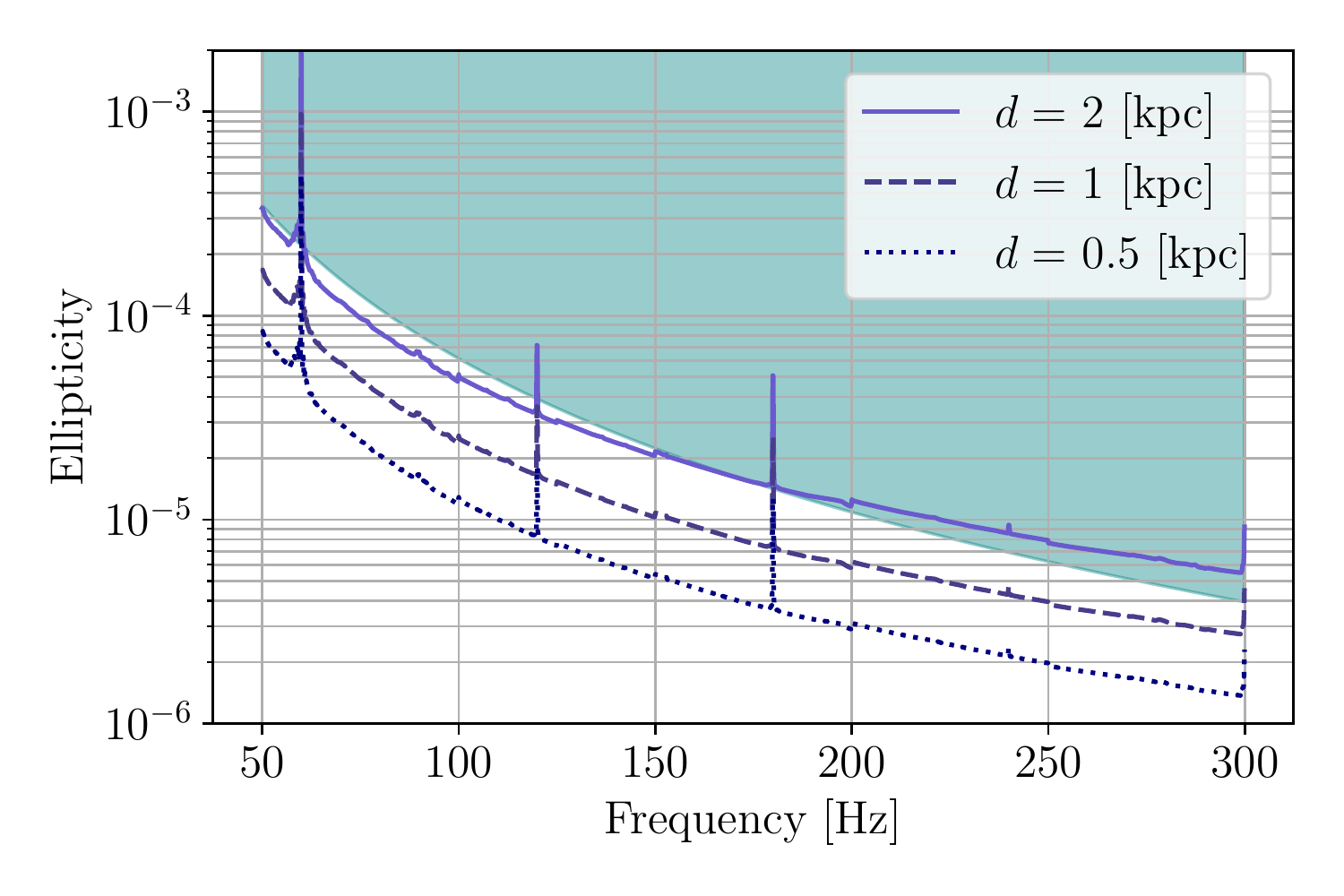}
    \caption{Maximum detectable ellipticity corresponding to the estimated sensitivity at different astrophysical
    ranges. Shaded regions are excluded due to the maximum spindown value probed by the search.
    Figure reproduced from~\protect\cite{O3a}.}
    \label{fig:ellipticity}
\end{figure}

Amplitude sensitivity estimations can be translated into information about the maximum allowed 
ellipticity of any probed source $\epsilon$ and the distance $d$ at which we would be able to detect it
if we assume the emission mechanism to be that of a non-axisymmetric NS
\begin{equation}
    h_0 = \frac{4 \pi² G}{c} \frac{I_{\textrm{z}}\epsilon}{d} f_{0}^2 \;,
\end{equation}
where $G$ is the gravitational constant, $c$ is the speed of light in vacuum and 
\mbox{$I_{\textrm{z}}=10^{38} \;\textrm{km} \cdot \textrm{m}^2$} is the canonical moment of
inertia of a NS around its rotation axis.
As shown in Fig.~\ref{fig:ellipticity}, we constrain NS ellipticities below $\epsilon = 10^{-5}$ 
for sources at $1\;\textrm{kpc}$ emitting at 150--300 Hz. Constraints below $\epsilon = 10^{-4}$ 
can be set for sources at $2\;\textrm{kpc}$ emitting at 75--150 Hz. These sensitivities approach the 
expected ellipticities of relativistic stars ($10^{-7} - 10^{-5}$ depending on the assumed equation of 
state). Neglecting the spindown of the source sets an upper limit on the maximum detectable ellipticity 
(shaded region in Fig.~\ref{fig:ellipticity}), although external influences such as accretion from a 
companion could allow us to probe within said region.


\newpage
\section*{Acknowledgements}
\begingroup
\scriptsize
The authors gratefully acknowledge the support of the United States
National Science Foundation (NSF) for the construction and operation of the
LIGO Laboratory and Advanced LIGO as well as the Science and Technology Facilities Council (STFC) of the
United Kingdom, the Max-Planck-Society (MPS), and the State of
Niedersachsen/Germany for support of the construction of Advanced LIGO 
and construction and operation of the GEO600 detector. 
Additional support for Advanced LIGO was provided by the Australian Research Council.
The authors gratefully acknowledge the Italian Istituto Nazionale di Fisica Nucleare (INFN),  
the French Centre National de la Recherche Scientifique (CNRS) and
the Netherlands Organization for Scientific Research, 
for the construction and operation of the Virgo detector
and the creation and support  of the EGO consortium. 
The authors also gratefully acknowledge research support from these agencies as well as by 
the Council of Scientific and Industrial Research of India, 
the Department of Science and Technology, India,
the Science \& Engineering Research Board (SERB), India,
the Ministry of Human Resource Development, India,
the Spanish Agencia Estatal de Investigaci\'on,
the Vicepresid\`encia i Conselleria d'Innovaci\'o, Recerca i Turisme, the Conselleria d'Educaci\'o i Universitat and the Direcci\'o General de Pol\'itica Universitaria i Recerca del Govern de les Illes Balears,
the Conselleria d'Innovaci\'o, Universitats, Ci\`encia i Societat Digital de la Generalitat Valenciana and
the CERCA Programme Generalitat de Catalunya, 
the Barcelona Supercomputing Center - Centro Nacional de Supercomputaci\'on, Spain,
the National Science Centre of Poland and the Foundation for Polish Science (FNP),
the Swiss National Science Foundation (SNSF),
the Russian Foundation for Basic Research, 
the Russian Science Foundation,
the European Commission,
the European Regional Development Funds (ERDF),
the Royal Society, 
the Scottish Funding Council, 
the Scottish Universities Physics Alliance, 
the Hungarian Scientific Research Fund (OTKA),
the French Lyon Institute of Origins (LIO),
the Belgian Fonds de la Recherche Scientifique (FRS-FNRS), 
Actions de Recherche Concert\'{e}es (ARC) and
Fonds Wetenschappelijk Onderzoek - Vlaanderen (FWO), Belgium,
the Paris \^{I}le-de-France Region, 
the National Research, Development and Innovation Office Hungary (NKFIH), 
the National Research Foundation of Korea,
the Natural Science and Engineering Research Council Canada,
Canadian Foundation for Innovation (CFI),
the Brazilian Ministry of Science, Technology, and Innovations,
the International Center for Theoretical Physics South American Institute for Fundamental Research (ICTP-SAIFR), 
the Research Grants Council of Hong Kong,
the National Natural Science Foundation of China (NSFC),
the Leverhulme Trust, 
the Research Corporation, 
the Ministry of Science and Technology (MOST), Taiwan,
the United States Department of Energy,
and
the Kavli Foundation.
The authors gratefully acknowledge the support of the NSF, STFC, INFN and CNRS for provision of computational resources.
This paper has been assigned document number LIGO-P2000298.

R.~T.~is supported by the Spanish Ministerio de Ciencia, Innovaci{\'o}n y Universidades 
(ref.~FPU 18/00694), European Union FEDER funds, the Ministry of Science, 
Innovation and Universities and the Spanish Agencia Estatal de Investigaci\'on 
grants PID2019-106416GB-I00/AEI/10.13039/501100011033, RED2018-102661-T, RED2018-102573-E, FPA2017-90687-REDC, 
Comunitat Autonoma de les Illes Balears through the Direcci\'o General de Pol\'itica Universitaria i Recerca 
with funds from the Tourist Stay Tax Law ITS 2017-006 (PRD2018/24), Generalitat Valenciana (PROMETEO/2019/071), 
EU COST Actions CA18108, CA17137, CA16214, and CA16104, and thankfully acknowledges the computer resources 
at CTE-Power and the technical support provided by Barcelona Supercomputing Center - Centro Nacional de 
Supercomputaci\'on through grant No. AECT-2019-3-0011 from the Red Espa\~nola de Supercomputaci\'on (RES).
\endgroup

\end{document}